\documentclass[intlimits,twoside,a4paper]{article}
\usepackage{amsmath,amssymb}
\usepackage{graphicx}
\usepackage{multirow}
\usepackage[english]{babel}
\usepackage{color}
\usepackage{caption}
\usepackage{subcaption}

\usepackage[T2A]{fontenc}
\usepackage[cp1251]{inputenc}

\usepackage[eqsecnum]{cmpj2}
\issue{2016}{19}{2}{23801}
\doinumber{10.5488/CMP.19.23801}

\title[Fluid of fused spheres]
{Fluid of fused spheres as a model for protein solution\thanks{We dedicate this contribution to our friend and coworker Professor A.D.J.~Haymet on occasion of his 60$^\text{th}$ birthday.}}
\author[M. Kastelic, Yu.V. Kalyuzhnyi, V. Vlachy]
{M. Kastelic\refaddr{label1}, Yu.V. Kalyuzhnyi\refaddr{label2}, V. Vlachy\refaddr{label1}}
\addresses{
\addr{label1} Faculty of Chemistry and Chemical Technology, University of
Ljubljana, Ve\v{c}na pot 113, 1000 Ljubljana, Slovenia
\addr{label2} Institute for Condensed Matter Physics of the National
Academy of Sciences of Ukraine, \\ 1 Svientsitskii St., 79011 Lviv, Ukraine
}

\authorcopyright{M. Kastelic, Yu.V. Kalyuzhnyi, V. Vlachy, 2016}
\date{Received November 17, 2015, in final form December 1, 2015}

\begin{document}

\maketitle

%%%%%%%%%%%%%%%%%%%%%%%%%%%%%%%%%%%%%%%%%%%%%%%%%%%%%%%%%%%%%%%%%%%%%%%

\begin{abstract}
\noindent In this work we examine thermodynamics of fluid with
``molecules'' represented by two fused hard spheres, decorated by
the attractive square-well sites. Interactions between these sites
are of short-range and cause association between the fused-sphere
particles. The model can be used to study the non-spherical (or
dimerized) proteins in solution. Thermodynamic quantities of the
system are calculated using a modification of Wertheim's
thermodynamic perturbation theory and the results compared with
new Monte Carlo simulations under isobaric-isothermal conditions.
In particular, we are interested in the liquid-liquid phase
separation in such systems. The model fluid serves to evaluate the
effect of the shape of the molecules, changing from spherical to
more elongated (two fused spheres) ones. The results indicate that
the effect of the non-spherical shape is to reduce the critical
density and temperature. This finding is consistent with
experimental observations for the antibodies of non-spherical
shape.
\pacs 87.15.A-, 87.15.ad, 87.15.km, 87.15.nr, 82.60.Lf, 64.70.Ja
\end{abstract}

%%%%%%%%%%%%%%%%%%%%%%%%%%%%%%%%%%%%%%%%%%%%%%%%%%%%%%%%%%%%%%%%%%%%%%%

\vspace{-3mm}

\section{Introduction}

Aggregation of proteins in solution is both desired and undesired.
It represents the first step in the downstream processing, i.e.,
salting out of the proteins for the purpose of cleaning and
application. It is also one of the intermediate steps in the
process of protein crystallization. The unwanted, pathological,
protein aggregation is known to cause several diseases. Very
importantly, bio-pharmaceutical drugs should be free of
aggregates, otherwise they may be harmful. To increase the
stability of protein in aqueous solutions is, therefore, an
important technical problem. For an excellent review of the
theoretical and experimental studies of protein solutions see
reference~\cite{Gunton2007}.

The class of proteins we are interested in here are the so-called
globular proteins. A typical representative of this class is
lysozyme, which was extensively studied both experimentally and
theoretically (see for example~\cite{Gunton2007}, Chapter~9).
Despite their complicated 3D structure, many properties of protein
solutions can be successfully described using relatively simple
models~\cite{Abramo2012,Lomakin1999,Ruppert2001,Rosch2007,
Gogelein2008,Carlsson2001,Kastelic2015}. Globular proteins are in
their native form (we assume that during the experimental
treatment they do not denature) most often pictured as perfectly
spherical objects. This na\" ive representation is in reality
never satisfied, it is used merely to simplify the calculations.
There is a large list of non-spherical proteins, for example the
shape of lysozyme mentioned above is ellipsoidal, including
antibodies, lactoferrin, and others, where more complex geometry
of particles would need to be taken into account to generate
realistic results. This is important because the interactions
leading to protein aggregation are directional and of short-range.

The shape of protein molecules influence their mutual interaction
directly and indirectly. For example, (i) depletion interaction is
largely dependent on the shape of
particles~\cite{Lekkerkerker2011}. (ii) Experimentally, it is
observed that many of proteins with roughly spherical shape
exhibit upper critical solution temperature at protein
concentration equal to
240~g/L~\cite{Gunton2007,Taratuta1990,Broide1991}. In contrast to
that, Y-shaped antibodies exhibit the shift toward much lower
values, way down to 100~g/L \cite{Mason2010,Wang2013}. (iii) The
hydrodynamic radii of the non-spherical objects are different,
therefore, their hydrodynamic and transport
coefficients~\cite{Yatsenko2007}, as well as, kinetic
parameters~\cite{Liu1995} are modified. It is also known that
classical nucleation theory has difficulties in describing the
crystallization of other than spherical (for example ellipsoidal)
particles~\cite{Gunton2007, Wheeler2012}.

Recently, we used a simple spherical model~\cite{Kastelic2015} to
analyze experimental results for the cloud-point temperatures in
aqueous protein solutions with added
salts~\cite{Taratuta1990,Broide1991}. We modelled the solution as
a one-component system; the protein molecules were represented as
perfect spheres, embedded in the continuum solvent composed of
water, buffer, and various simple salts. The attractive
short-range interactions between the proteins were due to the
square-well sites located on the surface of protein molecules. The
model was numerically evaluated using Wertheim's perturbation
theory~\cite{Wertheim1986_3,Wertheim1986_4,Chapman1988}. The
short-range and directional nature of the interactions among
proteins led to good agreement with the experimental data for the
liquid-liquid phase diagram in case of lysozyme and
$\gamma$-crystallin solutions~\cite{Taratuta1990,Broide1991}. With
knowledge of the experimental cloud-point temperature as a
function of composition of electrolyte present in the system, the
model gave predictions for the liquid-liquid coexistence curves,
the second virial coefficients, and other similar properties under
such experimental conditions.

One weakness of the model presented above was its simplified
geometry. Neither lysozyme nor other proteins are spherical, and
some of them for example, lactoferrin~\cite{Li2014} look more like
two fused spheres. The other weakness was that we assume for
protein molecules to exist in form of monomers, which is not true.
Even in very dilute solutions, proteins can be at least partially
dimerized. The purpose of the present work is to investigate how
the relaxing of these two basic assumptions influence the
liquid-liquid coexistence curve.

The models for the association of spherically symmetric particles
into dimer molecules are of considerable interest to science and
technology and have been actively studied earlier. Of particular
interest for us are the models where there is an inter-penetration
(``fusing'' of cores) of the spherical particles upon association
so that the bonding length $L$ is less than the core diameter
$\sigma $. The ``shielded sticky shell'' and the ``shielded sticky
point'' models of Stell and
co-workers~\cite{Cummings1984,Stell1989,Kalyuzhnyi1994} and their
extensions \cite{Pizio1994,Kalyuzhnyi1995,Duda1996}, belong to
this group of models and are the starting point for our work.
These types of the models were studied using regular
\cite{Cummings1984,Stell1989,Pizio1994} and multi-density
\cite{Wertheim1984_1,Wertheim1984_2,Kalyuzhnyi1993,Kalyuzhnyi1994,
Kalyuzhnyi1995,Duda1996} integral equation theories.

In the present study, we use spherical particles as building
blocks, which are fused together to form a new species. In this
way, we compose the molecule with arbitrary spacing $L$ between
the centers of spheres. Next, we decorate the surfaces of
fused-sphere molecules with the attractive short-range binding
sites, which may cause the association of the newly formed
molecules. Such an extension of the protein model follows from our
previous work~\cite{Kastelic2015}. Here, we wish to explore the
effects of the non-spherical shape on various thermodynamic
properties.

Different versions of the model of dimerizing particles,
represented by the tangentially bonded chain molecules, have been
studied earlier~\cite{Ghonazgi1993,Ghonazgi1994}. In this type of
the model, dimerization occurs due to square-well bonding site,
placed on the surface of one of the hard-sphere terminal monomer
of each chain. Theoretical description of the model was carried
out using first order thermodynamic perturbation theory (TPT1) of
Wertheim \cite{Wertheim1986_3,Wertheim1986_4}. There are two major
features of our model that set it apart from the models studied
earlier, i.e., (i) in our model the molecules are represented by
the two hard-sphere monomers fused at a distance $L$, which is
less than the contact distance $\sigma$ and (ii) the molecules
upon association can form a three-dimensional network. Due to the
former feature of the model, a straightforward application of
Wertheim's multi-density approach fails to produce accurate
results~\cite{Kalyuzhnyi1993,Duda2001_1,Duda2001_2}. In the
present work, we use a modified version of the TPT1, which takes
into account the change of the overall packing fraction of the
system due to the association
forces~\cite{Kalyuzhnyi1993,Urbic2007,Kalyuzhnyi2012}. The
accuracy of our modified TPT1 approach is checked by the newly
generated Monte Carlo simulation data.

%%%%%%%%%%%%%%%%%%%%%%%%%%%%%%%%%%%%%%%%%%%%%%%%%%%%%%%%%%%%%%%%%%%%%%%

%\newpage

\vspace{-2mm}

\section{Model, theory, and simulations}

\vspace{-1mm}

\subsection{Model}

\begin{figure}[!t]
\centering
\includegraphics[keepaspectratio=true,width=0.35\textwidth]{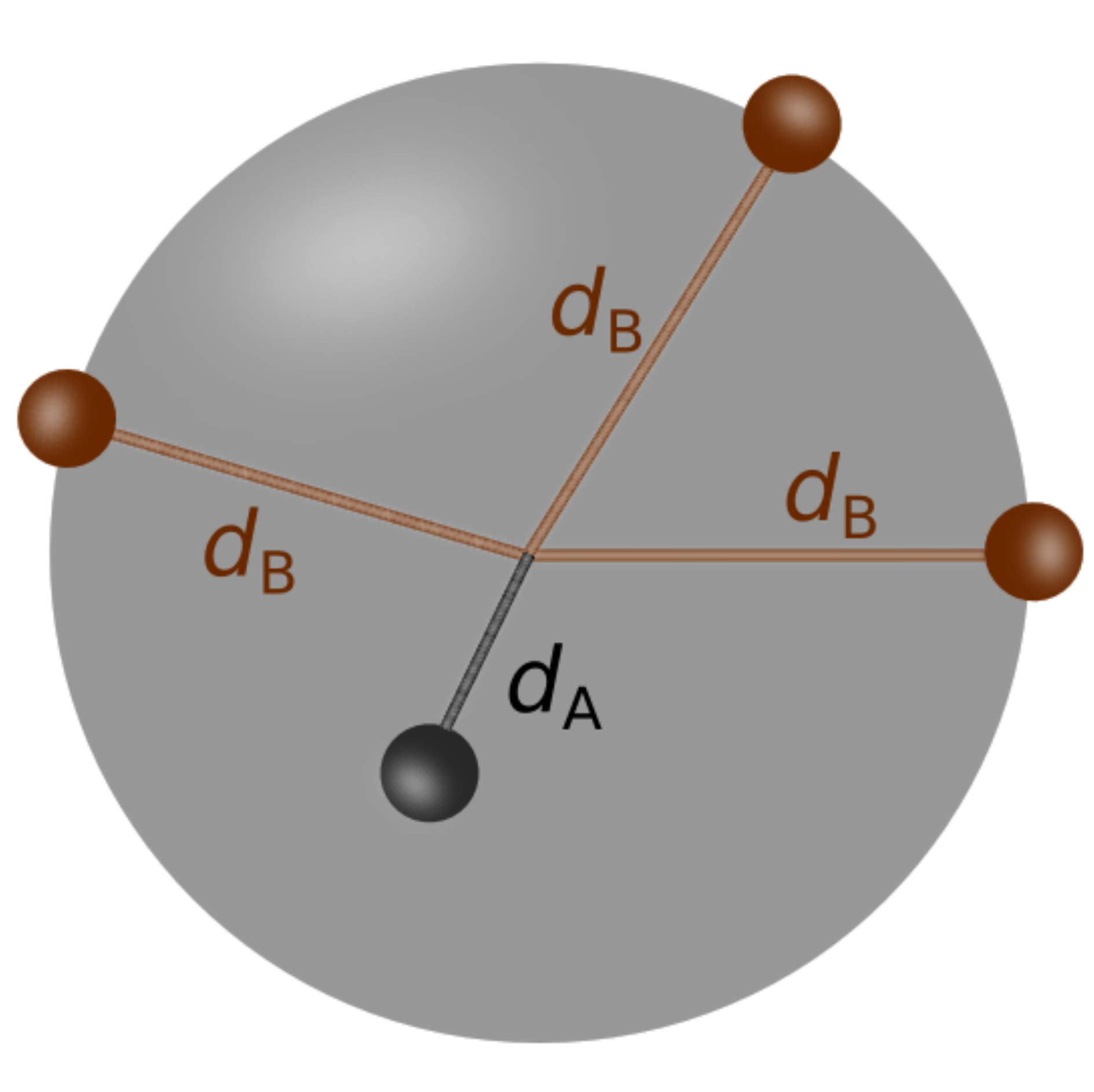}
\caption{(Color online). Spherical particles with diameter
$\sigma$ are capable of, due to the attraction among sites of type
A, penetrating to form fused sphere molecules. The cross
interactions A--B  are prohibited. In this figure,
\mbox{$K_\textrm{B}=3$}.} \label{figure1}
\end{figure}

We introduce a one component model of spherical particles,
decorated with additional binding sites of two different types, A
and B. The binding site A is placed within the sphere, with the
displacement \mbox{$d_\textrm{A}\leqslant\sigma/2$}, while an
arbitrary number $K_\textrm{B}$ of binding sites of type B is
located on the surface of the sphere (the displacement
\mbox{$d_\textrm{B}=\sigma/2$}). The model is visualized in
figure~\ref{figure1}. We consider a special case, where we exclude
the cross interactions among sites A and B.
The total pair potential is written as follows:
\begin{eqnarray}
u(\mathbf{{r}}) & = & u_\mathrm{R}(r)+\sum_{M=\textrm{A}}^\textrm{B}
u_{{MM}}(\mathbf{{x}}_{{MM}}), \label{Eq1}
\end{eqnarray}
where $u_\mathrm{R}$ is the pair potential for hard spheres, and
$u_\textrm{AA}$ and $u_\textrm{BB}$ are inter-particle site-site
potentials. The vector \mbox{$\mathbf{{r}}$ ($r=|\mathbf{{r}}|$)}
connects the centers of hard spheres, and $\mathbf{{x}}_{{MM}}$
denotes the inter-particle vector connecting two sites of the type
$M$. As mentioned above, $u_{{MM}}$ is the orientation dependent
square-well potential between the sites $M\in\{\text{A,B}\}$,
defined as follows:
\begin{eqnarray}
u_{MM}(\mathbf{{x}}_{MM}) & = &
\left\{\begin{array}{ll}
\varepsilon'_{MM}=-\varepsilon_{MM}-\xi_{MM}, &
\quad \text{for} \quad |\mathbf{{x}}_{MM}| < a_{MM}, \\
0, & \quad \text{for} \quad |\mathbf{{x}}_{MM}| \geqslant a_{MM}.
\end{array} \right. \label{Umm}
\end{eqnarray}
The site A causes inter-penetration of particles (see
figure~\ref{simulation}). Note that we need the term
$\xi_\textrm{AA}\rightarrow\infty$ to compensate for the hard
sphere repulsion. For the periphery sites B, we do not need such a
term, therefore $\xi_\textrm{BB}\rightarrow0$. To fuse hard cores
at separation $L$, we choose $d_\textrm{A}=L/2$ and take the limit
$\varepsilon_\textrm{AA}\rightarrow\infty$.

%%%%%%%%%%%%%%%%%%%%%%%%%%%%%%%%%%%%%%%%%%%%%%%%%%%%%%%%%%%%%%%%%%%%%%%

\subsection{Theory} \label{Theory_section}

An appropriate theoretical approach to be used is the first-order
Wertheim's thermodynamic perturbation theory
(TPT1)~\cite{Wertheim1986_3,Wertheim1986_4}. According to this theory, the
Helmholtz free energy of the system can be written as a sum
of several terms:
\begin{eqnarray}
A & = & A^\textrm{id}+A^\textrm{hs}+A^\textrm{A--A}+A^\textrm{B--B},
\end{eqnarray}
where \mbox{$A^\textrm{id}+A^\textrm{hs}=A_\textrm{R}$} is the
free energy of the reference system represented by the
hard-sphere system~\cite{Hansen_McDonnald2006} and
$A^\textrm{A--A}+A^\textrm{B--B}=A^\textrm{ass}$ is the
contribution due to A--A and B--B interactions. Following
Chapman et al.~\cite{Chapman1988}, we have:
\begin{eqnarray}
\frac{\beta(A-A_\textrm{R})}{N} & = &
\frac{\beta A^\textrm{A--A}}{N}+\frac{\beta A^\textrm{B--B}}{N},
\label{free_energy} \\
\frac{\beta A^\textrm{A--A}}{N} & = &
\ln{X_\mathrm{A}}-\frac{1}{2}X_\mathrm{A}+
\frac{1}{2}, \label{free_energyAA} \\
\frac{\beta A^\textrm{B--B}}{N} & = &
K_\textrm{B}\left(\ln{X_\mathrm{B}}-\frac{1}{2}X_\mathrm{B}+\frac{1}{2}
\right). \label{free_energyBB}
\end{eqnarray}
Here, $\beta=(k_\textrm{B}T)^{-1}$ and $k_\textrm{B}$ is
Boltzmann's constant as usual, $T$ is the absolute temperature,
and $N$ is the number of spheres. Further, $X_{M}$ defines the
average number fraction of particles, which are not bonded through
the binding site $M$. Parameters $X_\textrm{A}$ and $X_\textrm{B}$
are determined by the mass-action law~\cite{Chapman1988}
\begin{eqnarray}
X_\textrm{A} & = & \frac{1}{1+\rho\big(X_\textrm{A}\Delta_{\mathrm{AA}}+
K_\textrm{B}X_\textrm{B}\Delta_{\mathrm{AB}}\big)}, \\
X_\textrm{B} & = & \frac{1}{1+\rho\big(X_\textrm{A}\Delta_{\mathrm{BA}}+
K_\textrm{B}X_\textrm{B}\Delta_{\mathrm{BB}}\big)},
\label{mass-action}
\end{eqnarray}
where $\rho=N/V$ is the number density of spheres and
$\Delta_{MN}$ connects the pair distribution function of
hard spheres $g^{\mathrm{hs}}(r)$ (reference system) and the
binding potential for sites $M$ and $N$. The
corresponding $\Delta_{MN}$ parameters are:
\begin{eqnarray}
\Delta_{{MN}} & = & 4\pi
\int_{d_{M}+d_{N}}^{d_{M}+d_{N}+a_{MN}}
g^{\mathrm{hs}}(r)\bar{f}_{{MN}}(r)r^2\rd r \qquad \forall \;
{M,N}\in\{\text{A,B}\}. \label{Delta}
\end{eqnarray}
Expression for the solid-angle averaged Mayer function
\begin{eqnarray}
\bar{f}_{{MN}}(r)=\int\int
f_{{MN}}\big(\mathbf{{x}}_{{MN}}(\mathbf{
{r}})\big)\rd\Omega_{M}\rd\Omega_{N}
\end{eqnarray}
was initially derived by Wertheim~\cite{Wertheim1986} and further
generalized
here to be
\begin{eqnarray}
\bar{f}_{{MN}}(r) & = & \frac{\exp{\left(-\beta
\varepsilon'_{MM}\right)}-1}{24d_{M}d_{N}r}
(a_{MN}+d_{M}+d_{N}-r)^2(2a_{MN}-d_{M}
-d_{N}+r). \label{Mayer_function}
\end{eqnarray}
To suppress the cross interactions A--B, we set
$\Delta_{\mathrm{AB}}=\Delta_{\mathrm{BA}}=0$, which finally
yields two independent equations, written in a quadratic form
\begin{eqnarray}
\rho\Delta_{\mathrm{AA}}X_\textrm{A}^2+X_\textrm{A}-1 & = & 0, \label{MAL}
\\
\rho K_\textrm{B}\Delta_{\mathrm{BB}}X_\textrm{B}^2+X_\textrm{B}-1 & = &
0 \label{MAL1}.
\end{eqnarray}

%%%%%%%%%%%%%%%%%%%%%%%%%%%%%%%%%%%%%%%%%%%%%%%%%%%%%%%%%%%%%%%%%%%%%%%

\subsubsection{Association parameters $\Delta_\textrm{AA}$ and
$\Delta_\textrm{BB}$}

The association parameter $\Delta_{\mathrm{AA}}$ is related to
$X_\textrm{A}$ via equation (\ref{MAL}) and to the free energy
contribution due to A--A binding, by equation
(\ref{free_energyAA}). For the complete association limit, i.e.,
fusing of hard cores at separation $L$, no monomer spheres are
present, so $X_\textrm{A}=0$. We re-write the association
parameter $\Delta_{\mathrm{AA}}$ and introduce the cavity
correlation function $y^{\mathrm{hs}}(r)$ to obtain
\begin{eqnarray}
\Delta_{\mathrm{AA}} & = & 4\pi
\int_{2d_\textrm{A}}^{2d_\textrm{A}+a_\textrm{AA}}
y^{\mathrm{hs}}(r)e^{\mathrm{hs}}(r)\bar{f}_{\mathrm{AA}}(r)r^2\rd r,
\label{Delta1}
\end{eqnarray}
where $e^{\mathrm{hs}}(r)=\exp[-\beta u_\mathrm{R}(r)]$. Note
that, as already mentioned before, $2d_\textrm{A}=L$. By applying
the sticky limit approximation~\cite{Wertheim1986}, that is by
assuming the constant value of $y^{\mathrm{hs}}$ within the
integration domain, we obtain
\begin{eqnarray}
\Delta_{\mathrm{AA}} & = &
y^{\mathrm{hs}}(r=2d_\textrm{A})I_\textrm{AA},
\label{Delta_TS} \\
I_\textrm{AA} & = & 4\pi\int_{2d_\textrm{A}}^{2d_\textrm{A}+a_\textrm{AA}}
e^{\mathrm{hs}}(r)\bar{f}_{\mathrm{AA}}(r)r^2\rd r. \label{I_AA}
\end{eqnarray}
The integral given by equation~(\ref{I_AA}) is not used in further
calculations and, accordingly, it will not be considered in more
detail here.

Parameter $\Delta_\textrm{BB}$ determines the degree of
association of fused spheres and the free energy contribution due
to the B--B binding, see equations (\ref{free_energyBB}) and
(\ref{MAL1}). Notice that due to the association between $A$-type
of the sites, the packing fraction of fused spheres
$\eta_\textrm{eff}$ is different from the packing fraction
originally present (un-fused) hard spheres $\eta$. These fractions
are related as follows:
\begin{eqnarray}
\eta_\textrm{eff} & = & D(l^*)\eta, \label{D_par1} \\
D(l^*) & = &
\frac{1}{2}\bigg({1+\frac{3}{2}l^*-\frac{1}{2}{l^{*}}^3}\bigg),
\label{D_par2}
\end{eqnarray}
where $\eta=\pi\rho\sigma^3/6$ is the packing fraction of hard
spheres and $l^*=L/\sigma$ is reduced A--A bonding distance. Using
the sticky limit approximation~\cite{Wertheim1986} for
$\Delta_{\mathrm{BB}}$ [equation~(\ref{Delta})], we have:
\begin{eqnarray}
\Delta_{\mathrm{BB}} & = &
g^{\mathrm{hs}}(r=\sigma,\eta=\eta_\textrm{eff})I_\textrm{BB},
\label{Delta_EFF}\\
I_\textrm{BB} & = & 4\pi\int_{\sigma}^{\sigma+a_\textrm{BB}}
\bar{f}_{\mathrm{BB}}(r)r^2\rd r \label{DeltaBB}.
\end{eqnarray}
The integral in $I_\textrm{BB}$ can be evaluated analytically.
We have used the Carnahan-Starling approximation for the contact value of
$g^{\mathrm{hs}}$ at the effective packing fraction of fused
spheres $\eta_\textrm{eff}$
\begin{eqnarray}
g^{\mathrm{hs}}(r=\sigma,
\eta=\eta_\textrm{eff})=\frac{2-\eta_\textrm{eff}}{2(1-\eta_\textrm{eff})^3
}. \label{g_contact}
\end{eqnarray}

%%%%%%%%%%%%%%%%%%%%%%%%%%%%%%%%%%%%%%%%%%%%%%%%%%%%%%%%%%%%%%%%%%%%%%%

\subsubsection{Cavity correlation function $y^{\mathrm{hs}}$}

The last unknown quantity in equation~(\ref{Delta_TS}) is the
cavity correlation function of hard sphere fluid,
$y^{\mathrm{hs}}$. It is calculated by using the Tildesley-Streett
expression for pressure of the hard dumbbell
fluid~\cite{Tildesley1980}. By choosing $K_\textrm{B}=0$,
$d_\textrm{A}\leqslant\sigma/2$ and applying sticky limit
conditions, i.e.,
\mbox{$\varepsilon_\mathrm{AA}\rightarrow\infty$},
\mbox{$a_\mathrm{AA}\rightarrow0$} while keeping the integral in
equation (\ref{Delta1}) finite, our model reduces to the hard
dumbbell fluid. We modify the mass action law [equation
(\ref{MAL})], by inserting equation (\ref{Delta_TS}) with
$X_\textrm{A}=\rho_0/\rho$, where $\rho_0$ stands for the number
density of spheres, not bonded through binding site A (monomers).
The result represents a different form of equation (110) of
Wertheim's paper~\cite{Wertheim1986}
\begin{eqnarray}
\rho=\rho_0+\rho_0^2I_\textrm{AA}y^{\mathrm{hs}}(r=L). \label{quadratic}
\end{eqnarray}
Following Wertheim~\cite{Wertheim1986}, we get the expression for
the excess pressure in the form:
\begin{eqnarray}
\beta(P-P_\mathrm{R})=-\frac{1}{2}(\rho-\rho_0)\bigg\{1+\rho\frac{\partial
\ln[y^{\mathrm{hs}}(r=L)]}{\partial \rho}\bigg\}. \label{Wertheim}
\end{eqnarray}
We are now in position to obtain the cavity correlation
function $y^{\mathrm{hs}}$ of hard sphere system. We use the
Carnahan-Starling equation of state~\cite{Hansen_McDonnald2006}
for the reference system ($P_\mathrm{R}$) and the
Tildesley-Streett equation of state~\cite{Tildesley1980} for
the perturbed hard dumbbell system ($P$).
\begin{itemize}
\item Carnahan-Starling EOS:
%\end{itemize}
%
\begin{eqnarray}
\frac{\beta P_\text{R}}{\rho} & = &
\frac{1+\eta+\eta^2-\eta^3}{(1-\eta)^3} \label{CS}.
\end{eqnarray}
%
%\begin{itemize}
\item Tildesley-Streett EOS:
%\end{itemize}
%
\begin{eqnarray}
\frac{\beta P}{\rho_\mathrm{d}} & = &\!
\frac{1\!+\!(1\!+\!Ul^*\!+\!V{l^*}^3)\eta_\mathrm{eff}
\!+\!(1\!+\!Wl^*\!+\!X
{l^*}^3)\eta_\mathrm{eff} ^2\!-\!(1\!+\!Yl^*\!+\!Z{l^*}^3)\eta_\mathrm
{eff}^3}{(1\!-\!\eta_\mathrm{eff})^3} \label{TS},
\end{eqnarray}
\end{itemize}
where $\rho_\mathrm{d}=\rho/2$ is the number density of hard
dumbbells. The set of numerical parameters $U$, $V$, $W$, $X$,
$Y$, $Z$ is given in table~\ref{parameters}.

\begin{table}[!t]
\caption{Parameters in the Tildesley-Streett EOS~\cite{Tildesley1980}.\label{parameters}}
\vspace{2ex}
\centering
\begin{tabular}{|c|c|c|c|c|c|}
\hline\hline
$U$ & $V$ & $W$ & $X$ & $Y$ & $Z$ \\
\hline
0.37836 & 1.07860 & 1.30376 & 1.80010 & 2.39803 & 0.35700 \\
\hline\hline
\end{tabular}
\end{table}

\begin{figure}[!b]
\centering
\includegraphics[keepaspectratio=true,width=0.5\textwidth]{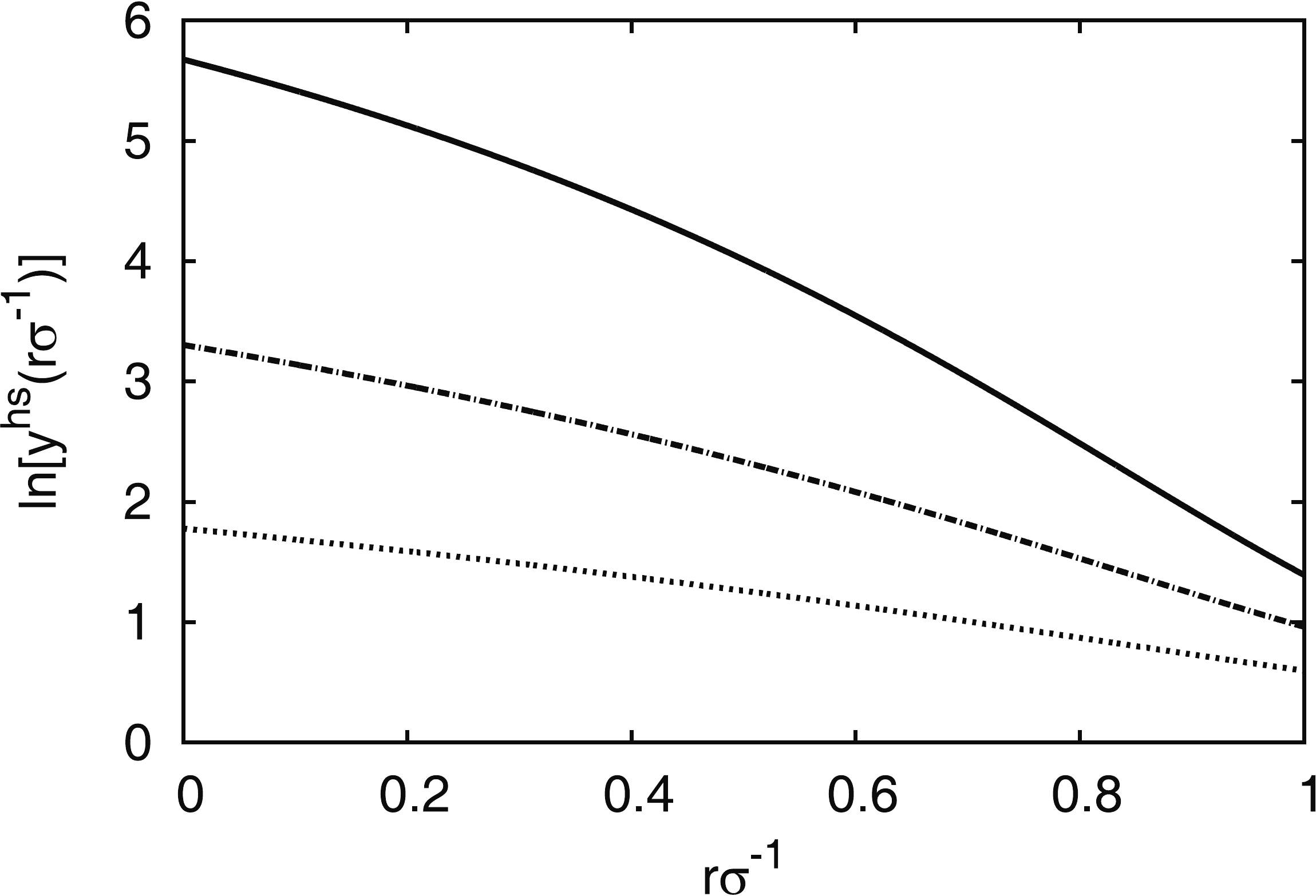}
\caption{Logarithm of the cavity distribution
function $y^{\mathrm{hs}}$ of hard spheres for $\rho\sigma^3$ is
equal to: 0.4 (dashed line), 0.6 (dashed-dotted line) and 0.8
(solid line). The limit of $y^{\mathrm{hs}}$,
$\lim_{r\rightarrow\sigma}y^{\mathrm{hs}}(r)$, coincides with
equation~(\ref{g_contact}) for the ``non-effective'' packing
fractions $\eta$.} \label{cavity_y}
\end{figure}

Within the framework of Wertheim's theory, we must set $\rho_0=0$
in equation (\ref{Wertheim}) to recover the fluid of hard dumbbell
particles (no monomers present). Next, we use equations
(\ref{D_par1}), (\ref{D_par2}), (\ref{Wertheim}) and equations of
state [(\ref{CS}) and (\ref{TS})] to obtain the derivative
\begin{eqnarray}
\rho\frac{\partial\ln[y^{\mathrm{hs}}(r=L)]}{\partial \rho} & = &
-\frac{\sum_{i=1}^{6}a_i\eta^i}{1+\sum_{i=1}^{6}b_i\eta^i}. \label{derive}
\end{eqnarray}
The set of equations which determine $a_i$ and $b_i$ [$D\equiv
D(l^*)$] are as follows:
\begin{eqnarray}
A & = & (1+Ul^*+V{l^*}^3)D, \\
B & = & (1+Wl^*+X{l^*}^3)D^2, \\
C & = & (1+Yl^*+Z{l^*}^3)D^3,
\end{eqnarray}
with the arrays
\begin{center}
\begin{tabular}{p{7cm} p{5.5cm}}
$a_1 = A+3D-8,$  &  $b_1 = -3(1 +D),$ \\
$a_2 = -3A+B+15D-3D^2+4,$  &  $b_2 = 3(1+3D+D^2),$ \\
$a_3 = 3A-3B-C-3D-15D^2+D^3,$  &  $b_3 = -(1+9D+9D^2+D^3),$\\
$a_4 = -A+3B+3C-3D+3D^2+5D^3,$  &  $b_4 = a_2D,$ \\
$a_5 = -B-3C+3D^2-D^3,$  &  $b_5 = a_1D^2,$ \\
$a_6 = C-D^3,$  &  $b_6 = D^3.$
\end{tabular}
\end{center}
It is obvious $\rho \,\partial \ln[y^{\mathrm{hs}}(r=L)]/\partial
\rho=\eta\,\partial\ln[y^{\mathrm{hs}}(r=L)]/\partial \eta$,
therefore equation~(\ref{derive}) can be easily integrated to
yield:
\begin{eqnarray}
\ln[y^{\mathrm{hs}}(r=L)] & = &
-\int_0^\eta\frac{\sum_{i=1}^{6}a_i t^{i-1}}{1+\sum_{i=1}^{6}b_i t^i}\rd t.
\label{integral}
\end{eqnarray}
The integral was checked to be non-singular for all investigated
$\eta$ and $l^*$ values. Numerical results for
$\ln[y^{\mathrm{hs}}(r)]$ are for a few fluid densities shown in
figure~\ref{cavity_y}.

%%%%%%%%%%%%%%%%%%%%%%%%%%%%%%%%%%%%%%%%%%%%%%%%%%%%%%%%%%%%%%%%%%%%%%%

\subsubsection{Other thermodynamic properties}

Next, we calculate the excess pressure
$P^\textrm{ass}=P-P_\textrm{R}$ and the excess chemical potential
$\mu^\textrm{ass}=\mu-\mu_\textrm{R}$ needed in phase diagram
calculations as also excess internal energy
$E^\textrm{ass}=E-E_\textrm{R}$, due to association. Starting with
the pressure, we have
\begin{equation}
\beta(P-P_\textrm{R}) =
\rho^2\frac{\partial\big[\beta(A-A_\textrm{R})/N\big]}
{\partial\rho}  =
\rho\sum_\textrm{M=A}^\textrm{B}\bigg(\frac{\partial\big[\beta(A-A_\textrm{
R})/N\big]}{\partial X_{M}} \bigg)\bigg(\eta\frac{\partial
X_{M}}{\partial \eta} \bigg).
\end{equation}
By inserting the appropriate derivatives from equations
(\ref{free_energy}), (\ref{MAL}), and (\ref{MAL1}), we get the
final expression for the excess pressure. The second term $B$ is
evaluated at $\eta_\textrm{eff}$, see equation~(\ref{g_contact}),
therefore upon differentiation we get an additional factor
$D(l^*)$
\begin{eqnarray}
\beta(P-P_\textrm{R}) & = & \beta P^\mathrm{AA}+\beta P^\mathrm{BB},
\label{P1} \\
\beta P^\mathrm{AA} & = &
-\frac{\rho}{2}(1-X_\textrm{A})\left\{1+\eta\frac{\partial
\ln[y^{\mathrm{hs}}(r=L)]}{\partial\eta}\right\}, \label{P2} \\
\beta P^\mathrm{BB} & = &
-\frac{\rho}{2}K_\textrm{B}(1-X_\textrm{B})\left\{1+D(l^*)\eta\frac{\partial
\ln[g^{\mathrm{hs}}(r=\sigma)]}{\partial\eta}\bigg|_{\substack{\eta=\eta_{
\textrm{eff}}}}\right\}. \label{P3}
\end{eqnarray}
The expression \mbox{$\eta\partial\ln[y^{\mathrm{hs}}(r=L)]/\partial \eta$}
is obtained from equation~(\ref{derive}), while the second derivative \linebreak
\mbox{$\partial\ln[g^{\mathrm{hs}}(r=L)]/\partial\eta\big|_{\substack{
\eta=\eta_{\textrm{eff}}}}$} is obtained analytically at
$\eta=\eta_\textrm{eff}$ from equation~(\ref{g_contact})
\begin{eqnarray}
\frac{\partial\ln[g^{\mathrm{hs}}(r=L)]}{\partial
\eta}\bigg|_{\substack{\eta=\eta_{\textrm{eff}}}} & = &
\frac{5-2\eta_\textrm{eff}}{(1-\eta_\textrm{eff})(2-\eta_\textrm{eff}
)}.
\end{eqnarray}

The excess chemical potential
$\mu^\textrm{ass}=\mu-\mu_\textrm{R}$ is obtained through the
relation
\begin{eqnarray}
\mu^\textrm{ass} & = &
\frac{A^\textrm{ass}}{N}+\frac{P^\textrm{ass}}{\rho}.
\end{eqnarray}
The logarithmic term $\ln X_\textrm{A}$ in
equation~(\ref{free_energyAA}) is divergent for the complete
association limit \mbox{($\Delta_{\mathrm{AA}}\gg1$)}, therefore
we re-write this term by using equation~(\ref{Delta_TS}) as
follows:
\begin{eqnarray}
\ln X_\textrm{A}=
\ln\bigg(\frac{-1+\sqrt{1+4\rho\Delta_{\mathrm{AA}}}}{2\rho\Delta_{
\mathrm{AA}}}\bigg){\approx}
\ln\bigg(\frac{\sqrt{4\rho\Delta_{\mathrm{AA}}}}{2\rho\Delta_{
\mathrm{AA}}}\bigg)=
-\frac{1}{2}\ln[\rho
y^{\textrm{hs}}(L)]-\frac{1}{2}\ln[I_\textrm{AA}(\beta)]. \label{additive}
\end{eqnarray}
The second term in equation~(\ref{additive}) is independent of
density and, accordingly, does not contribute to the pressure. The
expression for $P$ is the same as derived before [equations
(\ref{P1})--(\ref{P3})]. The equilibrium conditions require the
equality of chemical potential at a constant temperature (see
equations below), so the second term in equation~(\ref{additive})
cannot affect the coexistence curve. The equilibrium conditions
read:
\begin{eqnarray}
\mu(\rho_\textrm{I},T) & = & \mu(\rho_\textrm{II},T), \label{PD1} \\
P(\rho_\textrm{I},T) & = & P(\rho_\textrm{II},T), \label{PD2}
\end{eqnarray}
where $\rho_\textrm{I}$ and $\rho_\textrm{II}$ are the two
coexisting densities. At this step, the phase diagram can be
constructed by applying equations (\ref{PD1})--(\ref{PD2}) as it is in
more detail explained in the previous work~\cite{Kastelic2015}.

Another thermodynamic quantity is the excess internal
energy $E^\textrm{ass}=E-E_\textrm{R}$, obtained as
\begin{eqnarray}
\frac{E-E_\textrm{R}}{N} & = &
\frac{\partial\big[\beta(A-A_\textrm{R})/N\big]}{\partial\beta} =
\frac{E^\textrm{A--A}}{N}+\frac{E^\textrm{B--B}}{N}, \\
\frac{E^\textrm{A--A}}{N} & = & -\frac{\rho}{2}X_\textrm{A}^2
\frac{\partial\Delta_\textrm{AA}}{\partial\beta}, \\
\frac{E^\textrm{B--B}}{N} & = &
-\frac{\rho}{2}(K_\textrm{B}X_\textrm{B})^2
\frac{\partial\Delta_\textrm{BB}}{\partial\beta}.
\end{eqnarray}
Since ${E^\textrm{A--A}}/{N}$ is divergent, the only relevant part
is ${E^\textrm{B--B}}/{N}$. Derivative
$\partial\Delta_\textrm{BB}/\partial\beta$ is obtained
analytically from equations~(\ref{Delta_EFF})--(\ref{DeltaBB}),
since $g^{\mathrm{hs}}$ is $\beta$ independent. Thermodynamic
functions for the reference system of hard spheres, $\beta
A_\textrm{R}/N$, $\beta P_\textrm{R}$, $u_\textrm{R}$ and
$E_\textrm{R}/N$, can be found
elsewhere~\cite{Hansen_McDonnald2006}.

%%%%%%%%%%%%%%%%%%%%%%%%%%%%%%%%%%%%%%%%%%%%%%%%%%%%%%%%%%%%%%%%%%%%%%%
%%%%%%%%%%%%%%%%%%%%%%%%%%%%%%%%%%%%%%%%%%%%%%%%%%%%%%%%%%%%%%%%%%%%%%%

\subsection{$N,P,T$ Monte Carlo simulation}

\begin{figure}[!h]
\begin{center}
\includegraphics[keepaspectratio=true,width=0.48\textwidth]{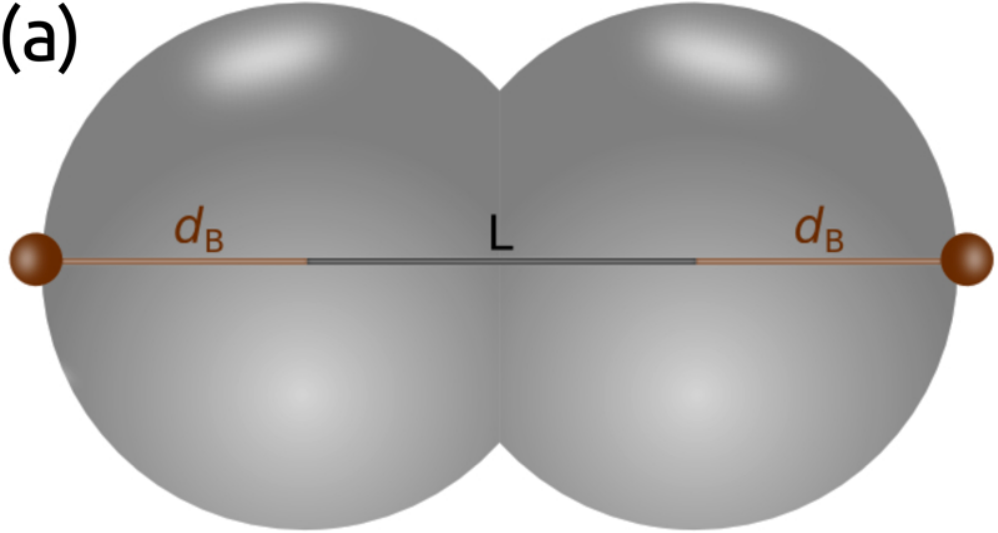}
\quad
\includegraphics[keepaspectratio=true,width=0.48\textwidth]{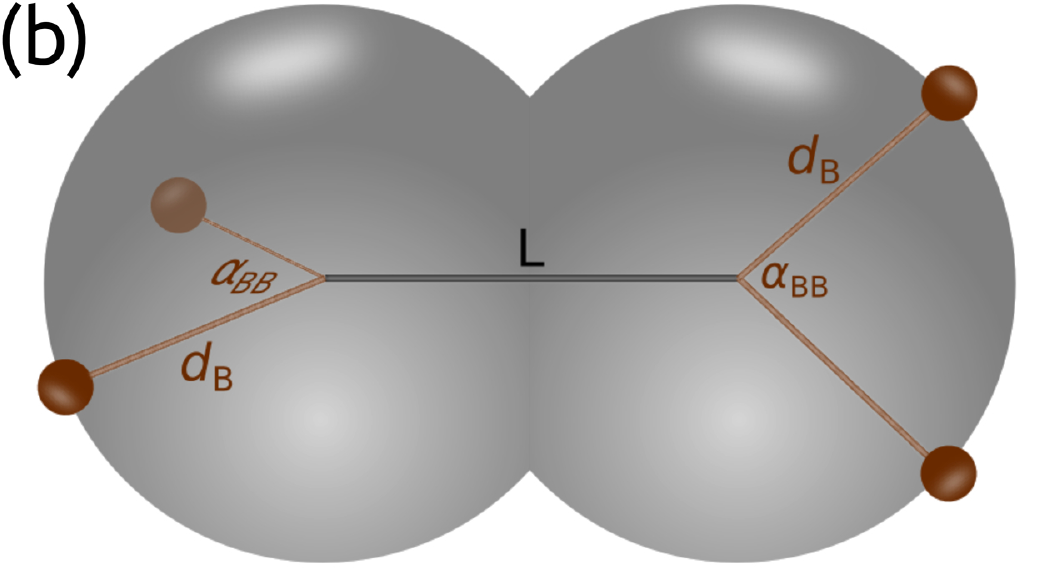}
\end{center}
\caption{(Color online). Different molecules in $N,P,T$ Monte Carlo simulations:
fixed binding sites on the opposite poles, $K_\textrm{B}=1$ (a), and more
complex geometry with two binding
sites on each sphere, $K_\textrm{B}=2$ (b). In the last example we set
$\alpha_\textrm{BB}=\pi/2$ with perpendicular orientation of lines,
connecting sites B on each sphere. Center-to-center separation $L$
($l^*=L/\sigma$) and displacement distance $d_\textrm{B}=\sigma/2$ were fixed.}
\label{simulation}
\end{figure}

To validate the accuracy of the modified TPT1 approach, we performed Monte
Carlo computer simulations in the $N,P,T$
ensemble~\cite{Frenkel1996}. We assumed fused spheres with
one and two binding sites on each sphere, where the prescribed
arrangement of sites was preserved during the simulation. Simulated
molecules are schematically shown in figure~\ref{simulation}.
We adopted the sampling method suggested by Tildesley and
Streett~\cite{Streett1976}, where a single displacement parameter
was needed to describe the translation and rotation of fused
spheres. The simulation box contained 250 fused spheres
(molecules), which is equivalent to 500 penetrating (original,
un-fused) spheres. We defined the \emph{cycle} with 250 attempts
to move the object and by 1 attempt to change the volume box.
Next, we defined the \emph{block} to be equal to $5\times10^4$
cycles. Initially, we performed 1 block, to equilibrate the
system, while 4 independent blocks were needed to calculate
thermodynamic properties via the block averaging. Simulations were
performed for three $l^*$ values: 0.2, 0.6, 1.0, and four
different pressures $Pk_\text{B}T/\sigma^3$: 0.5, 1.0, 2.0, and
4.0, for each model object visualized in figure~\ref{simulation}.
The acceptance rate of trial configurations was between 0.2 and
0.6.

%

%%%%%%%%%%%%%%%%%%%%%%%%%%%%%%%%%%%%%%%%%%%%%%%%%%%%%%%%%%%%%%%%%%%%%%%

%Results and discussion
\section{Results and discussion}

%%%%%%%%%%%%%%%%%%%%%%%%%%%%%%%%%%%%%%%%%%%%%%%%%%%%%%%%%%%%%%%%%%%%%%%

\subsection{Thermodynamic properties: Theory against Monte Carlo
simulations}

\begin{figure}[!t]
\begin{center}
%\begin{minipage}{0.5\columnwidth}
%\centering
\includegraphics[keepaspectratio=true,width=0.48\textwidth]{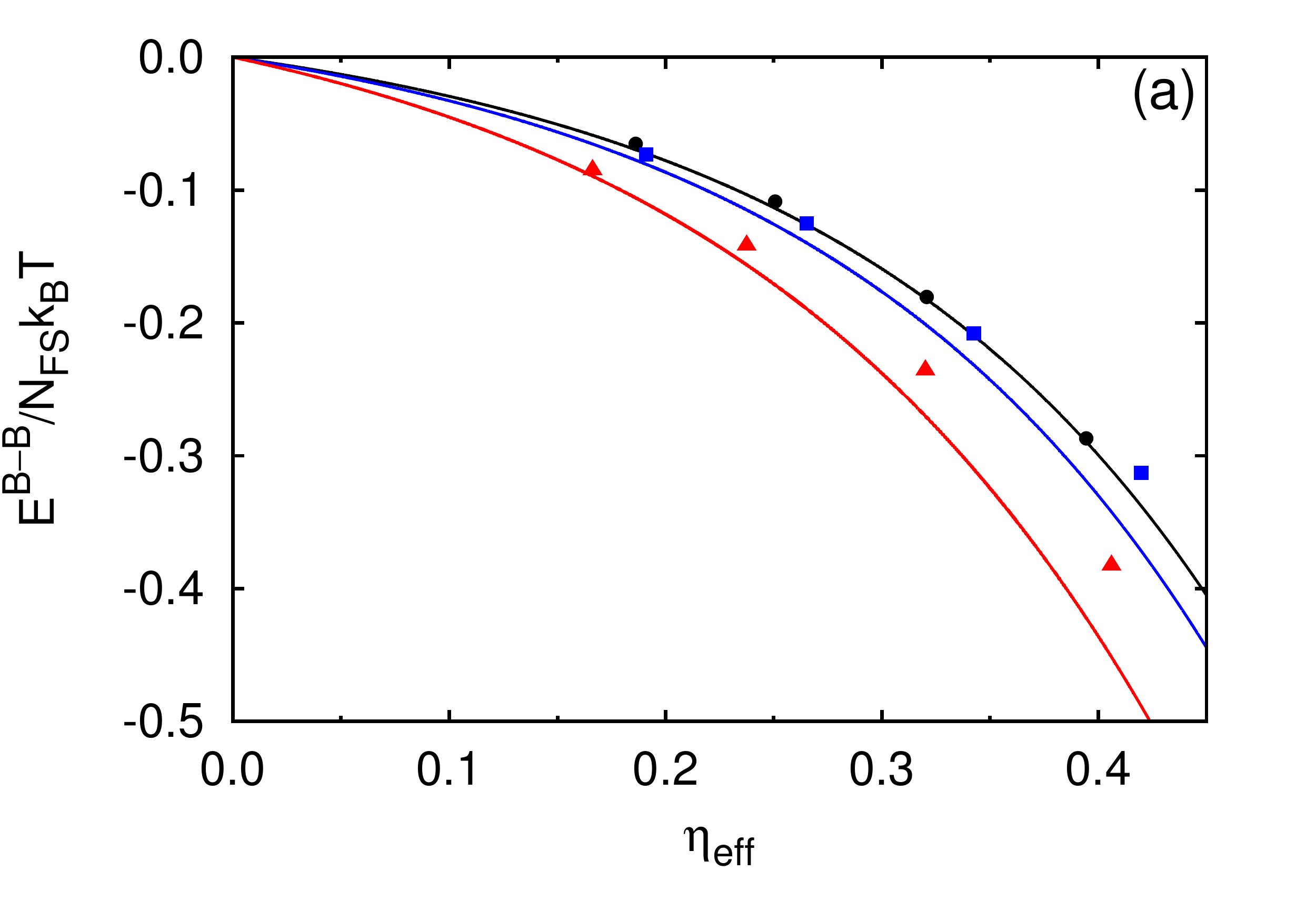}\quad
%\end{minipage}
%\hfill
%\begin{minipage}{0.5\columnwidth}
%\centering
\includegraphics[keepaspectratio=true,width=0.48\textwidth]{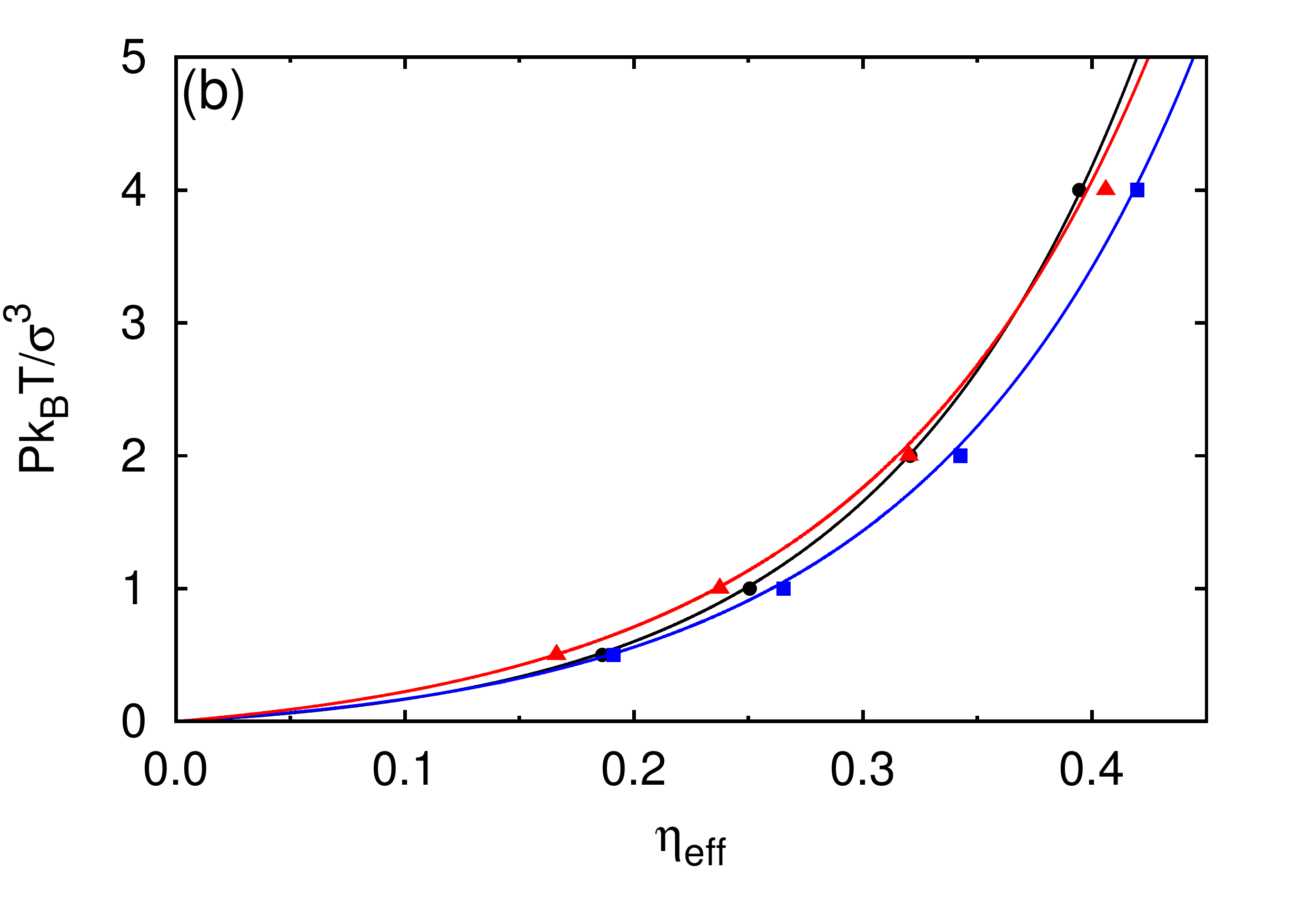} \\
%\end{minipage}
%\begin{minipage}{0.5\columnwidth}
%\centering
\includegraphics[keepaspectratio=true,width=0.48\textwidth]{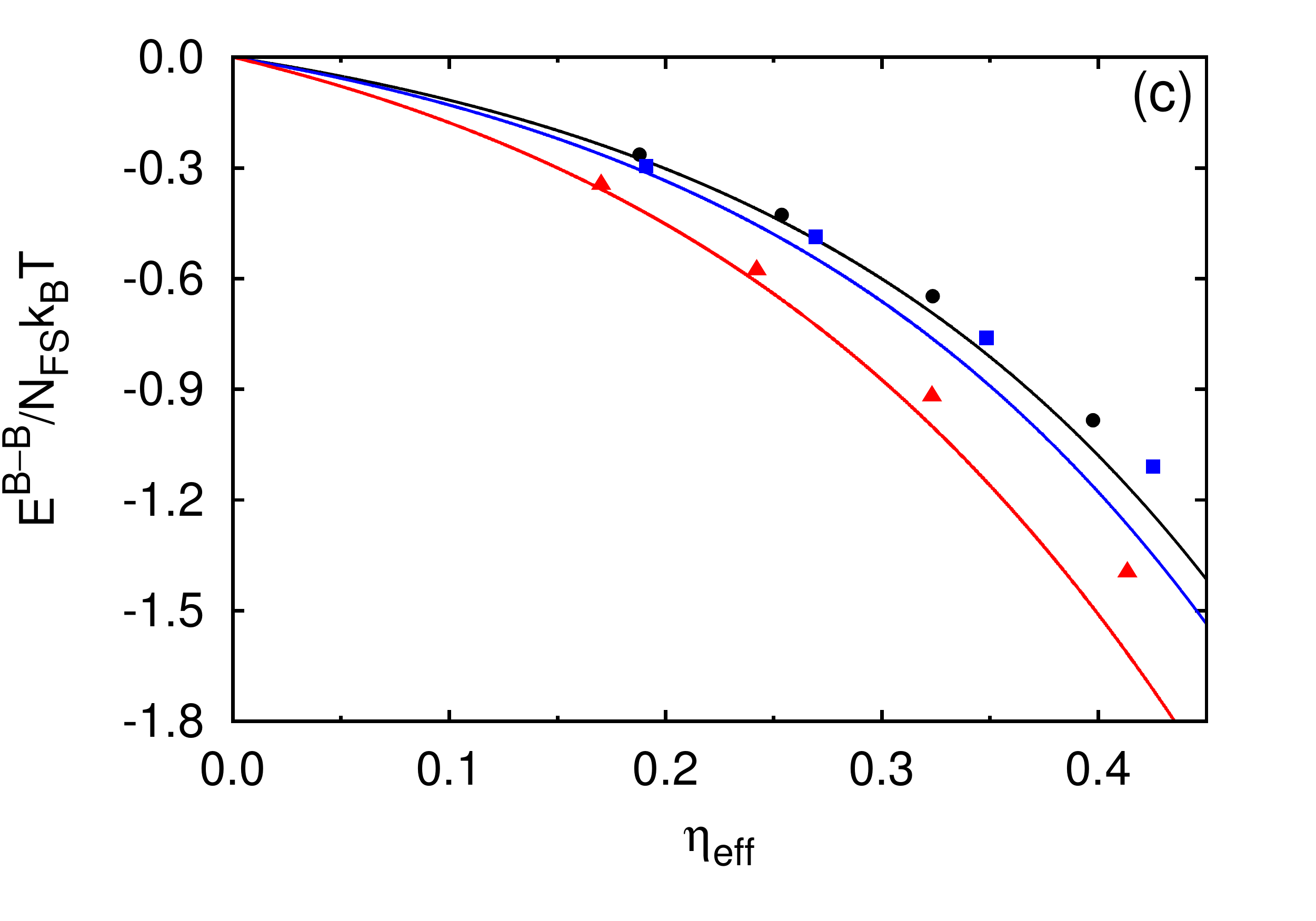}\quad
%\end{minipage}
%\hfill
%\begin{minipage}{0.5\columnwidth}
%\centering
\includegraphics[keepaspectratio=true,width=0.48\textwidth]{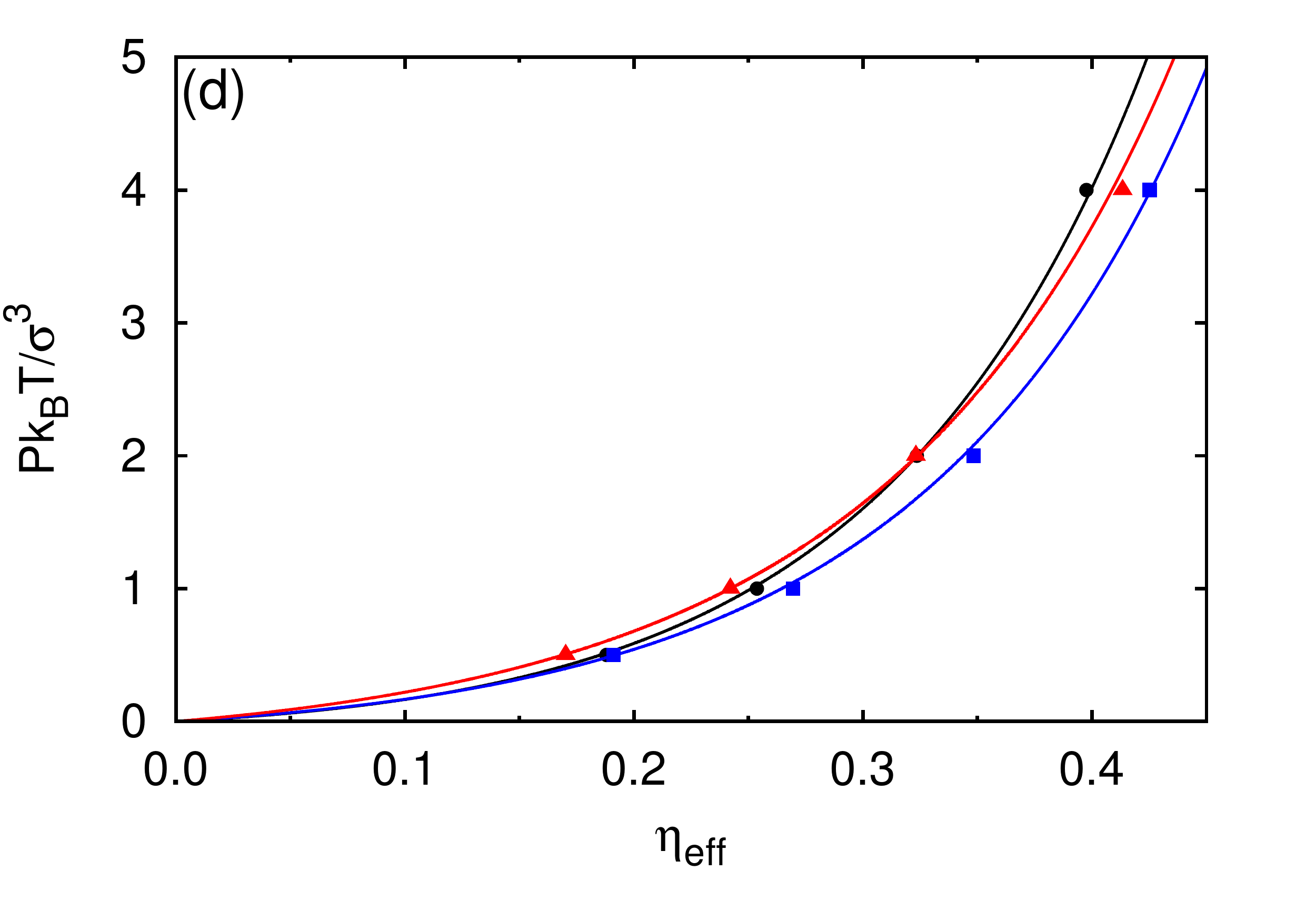}
%\end{minipage}
\end{center}
\vspace{-6mm}
\caption{(Color online). Pressure $P$ and association energy
$E^\textrm{B--B}$ per pair of fused spheres, thus
$N_\textrm{FS}=N/2$. The calculations
are presented by lines and the corresponding simulation results
by symbols. We studied three different $l^*$ values: 0.2
(red, \textcolor{red}{\large$\blacktriangle$}), 0.6
(blue, \textcolor{blue}{$\blacksquare$}) and 1.0
(black, \textcolor{black}{\Large$\bullet$}). Panels, (a)
and (b), belong to $K_\textrm{B}=1$ and the panels, (c) and
(d), to the case with $K_\textrm{B}=2$. Calculation apply to
$T^*=k_\text{B}T/\varepsilon=1$; pair potential parameters are
listed in table~\ref{PP_parameter}, note that
$L=2d_\textrm{A}$. Uncertainties of simulation are within the size of
symbols.} \label{T_comparison}
\end{figure}

To test the accuracy of TPT1 we performed $N,P,T$ Monte Carlo
simulations for values of $K_\textrm{B}$ equal to 1 and 2 (three $l^*$
values for each $K_\textrm{B}$). We chose to compare the pressure $P$ and
the internal energy due to B--B binding,
$E^\textrm{B--B}$. Since we used the complete A--A association
limit within TPT1, the latter quantity, $E^\textrm{B--B}$, was the one
that could be directly compared to computer
simulations. We fixed the temperature $T^*=k_\text{B}T/\varepsilon=1$, while
the pair potential characteristics are given in
table~\ref{PP_parameter}.
\begin{table}[htb!]
\caption{Pair potential parameters used for testing TPT1 againt
simulations.\label{PP_parameter}}
\vspace{2ex}
\centering
\begin{tabular}{cc}
\hline
\hline
$a_\textrm{BB}$: & 0.1$\sigma$ \\
$\varepsilon_\textrm{BB}$: & 5.0$\varepsilon$ \\
$d_\textrm{B}$: & 0.5$\sigma$ \\
\hline
\hline
\end{tabular}
\end{table}
The comparison between the theory and simulations is presented in
figure~\ref{T_comparison}. We found very good agreement for the
pressure, while the theoretical predictions for $E^\textrm{B--B}$
were less accurate. In case of $l^*=1$ we obtained very good
agreement for $K_\textrm{B}=1$ and fair agreement for
$K_\textrm{B}=2$ (black lines and corresponding symbols in
$E^\textrm{B--B}/N_\textrm{FS}k_\text{B}T$ sub-figures). If we
reduced the $l^*$ values (blue and red lines, symbols), the
deviations became larger, though the qualitative picture remained
correct. Deviations at low $l^*$ could be caused by the facts
that: (i) fusing of two spheres at small $l^*$ is not a small
perturbation regarding the reference system of hard spheres, and
(ii) the arrangement of binding sites B is fixed during the
simulation, which is not the case in TPT1, where the orientation
average over all geometries was assumed.

%%%%%%%%%%%%%%%%%%%%%%%%%%%%%%%%%%%%%%%%%%%%%%%%%%%%%%%%%%%%%%%%%%%%%%%

\subsection{Effect of protein's shape on the liquid-liquid phase diagram}

To illustrate the influence of protein shape on the liquid-liquid
phase behavior, we compared the phase diagrams for two versions of
the model: model (I) of two fused hard spheres and the model (II)
of equivalent hard sphere, which was defined as a limiting case of
(I), when $L\rightarrow 0$ and $\sigma\rightarrow d_\textrm{eqv}$.
The latter was chosen in such a way, that the volume of two fused
spheres in case (I) was equal to that of the equivalent sphere
(II): $d_\textrm{eqv}=\sigma\sqrt[3]{1+3l^*/2-l{^*}^3/2}$. Example
(II) might be interpreted as the usual hard sphere model of
diameter $d_\textrm{eqv}$, with $2K_\textrm{B}$ of sites B, i.e.,
the same number as on the two fused spheres. In such an
interpretation, the A--A contributions to the physical properties
can be neglected. Describing the aggregation of fused spheres of
diameter $d_\textrm{eqv}$ within the limiting conditions
$l^*\rightarrow 0$ and $D(l^*)\rightarrow \frac{1}{2}$, where
$\eta_\textrm{eff}=\eta/2=\pi\rho
d_\textrm{eqv}^3/12=\pi\rho_\textrm{d} d_\textrm{eqv}^3/6$, led us
to the model examined in reference~\cite{Kastelic2015}. Other
parameters and relations between examples (I) and (II) are listed
in table~\ref{parameters_shape_effect}.

\begin{table}[!t]
\caption{Pair potential parameters and relations used in
investigation of effects of protein's shape.\label{parameters_shape_effect} }
\vspace{2ex}
\centering
\begin{tabular}{c|c|c}
\hline\hline
 & two fused spheres at $L$ (I) & equivalent sphere (II) \\
\hline
diameter: & $\sigma$ & $d_\textrm{eqv}=\sigma\sqrt[3]{1+1.5l^*-0.5l{^*}^3}$
\\
$\eta_\textrm{eff}:$ & $ D(l^*)\eta$ & $\pi\rho_\textrm{d}
d_\textrm{eqv}^3/6$ \\
$\varepsilon_\textrm{BB}$: &
$\varepsilon$ & $\varepsilon$ \\
$d_\textrm{B}$: & $\sigma/2$ & $d_\textrm{eqv}/2$ \\
$a_\textrm{BB}$: & 0.1$\sigma$ & 0.1$\sigma$ \\
number of sites B: & $K_\textrm{B}$ (per building block) & $2K_\textrm{B}$
\\
\hline
\hline
\end{tabular}
\end{table}

\begin{figure}[!b]
\begin{center}
%\begin{minipage}{0.5\columnwidth}
%\centering
\includegraphics[keepaspectratio=true,width=0.48\textwidth]{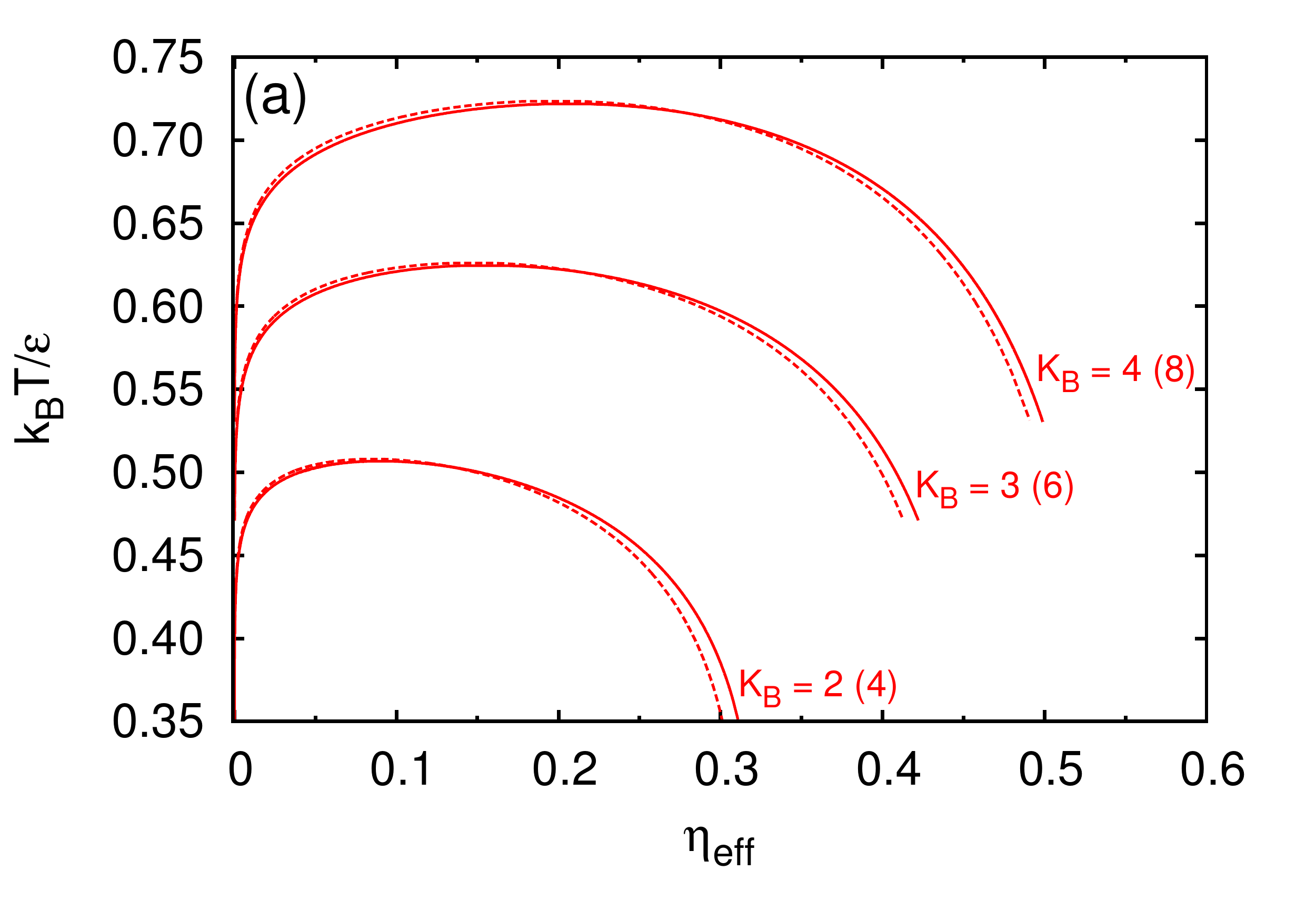} \quad
%\end{minipage}
%\hfill
%\begin{minipage}{0.5\columnwidth}
%\centering
\includegraphics[keepaspectratio=true,width=0.48\textwidth]{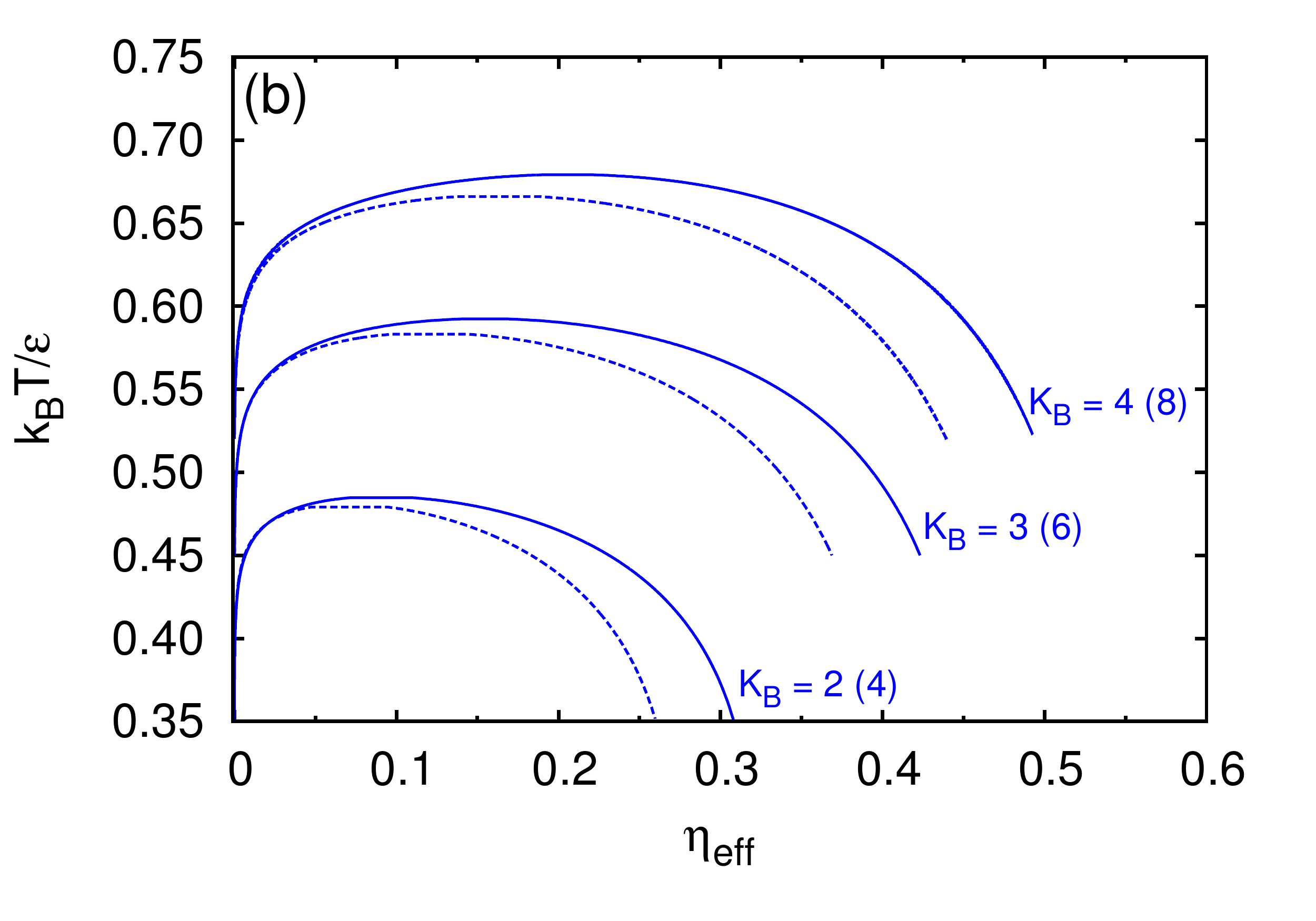}  \\
%\end{minipage}
%\begin{minipage}{1.0\columnwidth}
%\centering
\includegraphics[keepaspectratio=true,width=0.48\textwidth]{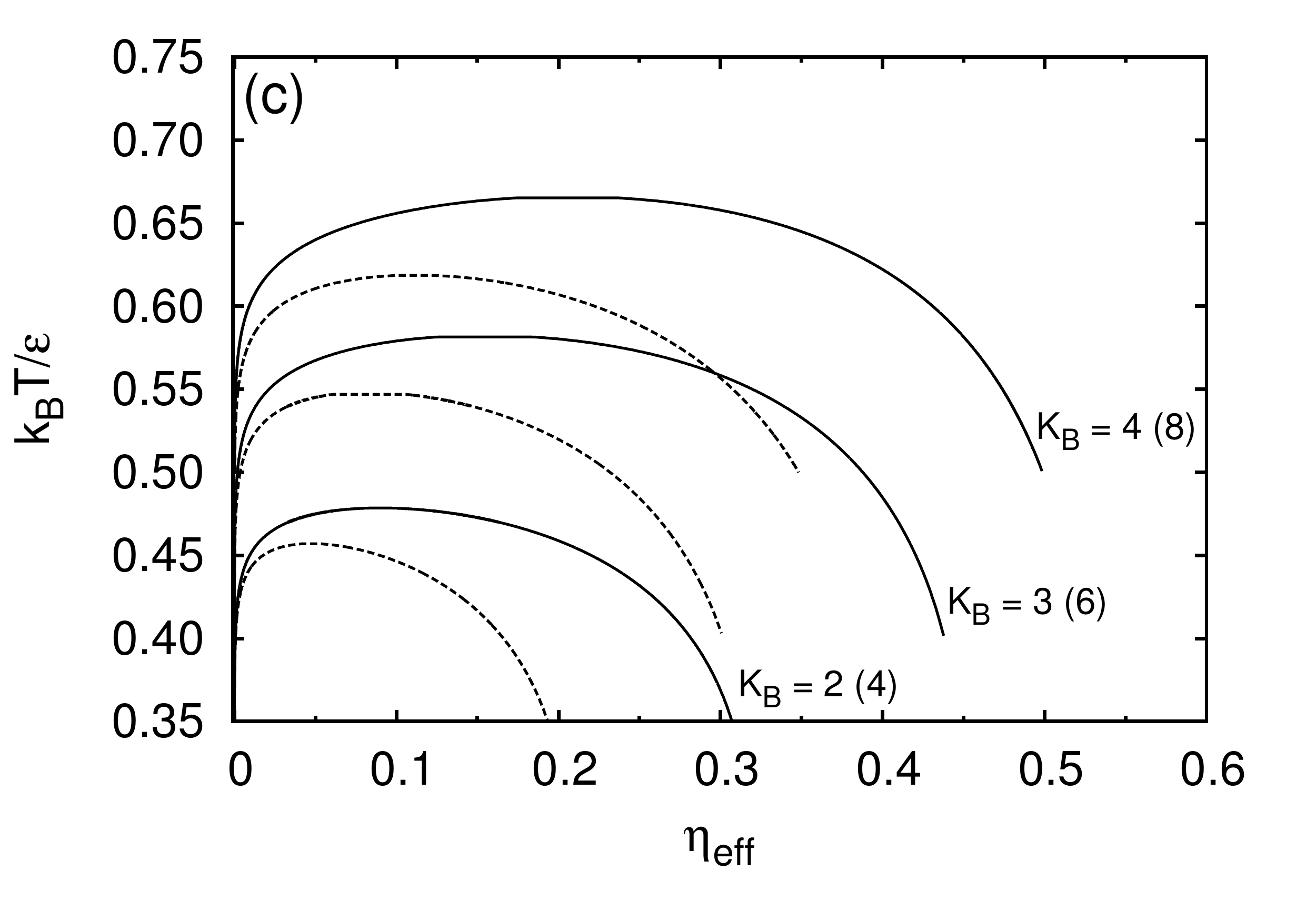}
%\end{minipage}
\end{center}
\vspace{-6mm}
\caption{(Color online). Phase diagrams for different $l^*$
values: 0.2 (a), 0.6 (b), and 1.0 (c), color notation is the same as
in figure~\ref{T_comparison}. Model of two fused spheres (I) is
denoted by dashed lines and the limiting model of equivalent
sphere (II) by solid lines. The results apply to three
different $K_\textrm{B}$ values, written without (I) and within
brackets (II), respectively.} \label{shape_effect}
\end{figure}

In figure~\ref{shape_effect}, we show phase diagrams for the
variants (I) and (II) described above, at three $l^*$ values and
for different numbers of sites B.
As observed before~\cite{Bianchi2006}, an increase of the number
of sites B shifts the critical density toward higher values. What
is more interesting here is the effect of the separation distance
parameter $l^*$ on the phase behavior. For a sufficiently small
$l^*$, i.e., $l^*=0.2$~--- see figure~\ref{shape_effect}~(a), the
difference between the phase diagrams for two versions of the
model becomes negligible, regardless of the $K_\textrm{B}$ value.
If centers of fused spheres are located at larger distance,
$l^*=0.6$~--- see figure~\ref{shape_effect}~(b), the difference
becomes more pronounced: both critical temperature and density are
lowered. Deviations become the strongest for the limiting example
of two spheres fused in contact, that is for $l^*=1.0$, cf.
figure~\ref{shape_effect}~(c). In this case, the larger number of
sites (larger area available for interaction) additionally affects
the liquid-liquid phase diagram. The shift toward lower critical
densities (or packing fractions) is consistent with experimental
studies of the Y-shaped antibodies~\cite{Mason2010,Wang2013}.

%%%%%%%%%%%%%%%%%%%%%%%%%%%%%%%%%%%%%%%%%%%%%%%%%%%%%%%%%%%%%%%%%%%%%%%
%%%%%%%%%%%%%%%%%%%%%%%%%%%%%%%%%%%%%%%%%%%%%%%%%%%%%%%%%%%%%%%%%%%%%%%
%%%%%%%%%%%%%%%%%%%%%%%%%%%%%%%%%%%%%%%%%%%%%%%%%%%%%%%%%%%%%%%%%%%%%%%

%Conclusions
\section{Conclusions}

Proteins come in many shapes, from ellipsoidal to Y-like and are
never perfectly spherical as treated by most theoretical models.
Further, even in dilute solutions they have a tendency to form
dimers and can be represented by two fused spheres. For dense
systems close to precipitation, the actual geometry of the protein
molecules is important; the inter-particle interactions are
directional and of short-range. In the present study, we modify
the first-order thermodynamic perturbation theory for associating
fluids to be applicable to the models allowing hard-sphere
particles to inter-penetrate. These particles can further
aggregate. We confront theoretical predictions for thermodynamic
properties of the proposed model with predictions of the
corresponding Monte Carlo simulations. We obtain an excellent
agreement for the pressure and fair agreement for the excess
internal energy. Next, we use this model to predict the
liquid-liquid phase diagram for protein solutions. We are
interested in the effects of protein shape on the phase
coexistence curve. We show that the fused hard-sphere model
reduces the critical density of the system in comparison with the
same quantity calculated for the hard-sphere model. This finding
is consistent with experimental observations for Y-shaped
antibodies. Using the cloud-point temperature measurements, we
currently investigate the influence of various salts on the
stability of lactoferrin solutions in water. The latter protein
has a shape of two fused spheres, and the hard-sphere model is not
a good representation of it. Theoretical approach developed in
this paper will be used to analyze experimental data for
lactoferrin and some other proteins of non-spherical geometry.

%%%%%%%%%%%%%%%%%%%%%%%%%%%%%%%%%%%%%%%%%%%%%%%%%%%%%%%%%%%%%%%%%%%%%%%
%%%%%%%%%%%%%%%%%%%%%%%%%%%%%%%%%%%%%%%%%%%%%%%%%%%%%%%%%%%%%%%%%%%%%%%
%%%%%%%%%%%%%%%%%%%%%%%%%%%%%%%%%%%%%%%%%%%%%%%%%%%%%%%%%%%%%%%%%%%%%%%

%Acknowledgments
\section*{Acknowledgements}
This study was supported by the Slovenian Research Agency fund
(P1-0201), NIH research grant (GM063592), and by the Young
Researchers Program (M.K.) of the Republic of Slovenia.

%%%%%%%%%%%%%%%%%%%%%%%%%%%%%%%%%%%%%%%%%%%%%%%%%%%%%%%%%%%%%%%%%%%%%%%
%%%%%%%%%%%%%%%%%%%%%%%%%%%%%%%%%%%%%%%%%%%%%%%%%%%%%%%%%%%%%%%%%%%%%%%
%%%%%%%%%%%%%%%%%%%%%%%%%%%%%%%%%%%%%%%%%%%%%%%%%%%%%%%%%%%%%%%%%%%%%%%

\ukrainianpart

\title{Плин зі спаяних сфер як модель розчину протеїнів
}
\author{М. Кастеліч\refaddr{label1}, Ю.В. Калюжний\refaddr{label2}, В.~Влахі\refaddr{label1}}
\addresses{
\addr{label1} Факультет хімії і хімічної технології, Університет Любляни, вул. Вечна, 113, 1000 Любляна, Словенія
\addr{label2} Інститут фізики конденсованих систем НАН України, вул. І. Свєнціцького, 1,
79011 Львів, Україна
}

\makeukrtitle

\begin{abstract}
\tolerance=3000%
В цій роботі ми досліджуємо термодинаміку плину з ``молекулами'', представленими двома спаяними твердими сферами, які декоровані вузлами з притягувальними потенціалами типу квадратної ями. Взаємодія між цими вузлами є короткодіюча і спричиняє асоціацію між частинками спаяних сфер. Модель може бути використана для дослідження несферичних (чи димеризованих) протеїнів у розчині. Термодинамічні величини системи розраховуються за допомогою модифікації термодинамічної теорії збурень Вертгайма, і результати порівнюються з новими симуляціями методом Монте Карло при ізобарично-ізотермічних умовах. Зокрема, нас цікавить фазове розшарування рідина-рідина в таких системах. Модельний плин використовується для оцінки ефекту форми молекул, що змінюється від сферичної до більш видовженої (дві спаяні сфери). Результати вказують, що ефект несферичної форми має зменшувати критичну густину і температуру. Це узгоджується з експериментальними спостереженнями для антитіл із несферичною формою.

\keywords несферичні протеїни, перехід рідина-рідина, напрямляюча сила, агрегація, термодинамічна
теорія збурень

\end{abstract}

\end{document}